\documentclass[12pt]{article}
\def\Slash#1{#1\kern-0.55em\raise.05ex\hbox{/}}
\usepackage{amsmath,amssymb,amsfonts,amscd,bbm,cancel,graphicx,hyperref}
\begin{document}


\thispagestyle{empty}
\renewcommand{\thefootnote}{\fnsymbol{footnote}}
\setcounter{footnote}{1}

\vspace*{-0.5cm}

\centerline{\Large\bf Weakly Coupled Discretized Gravity}

\vspace*{18mm}

\centerline{\large\bf
Gerhart Seidl\footnote{E-mail: \texttt{seidl@physik.uni-wuerzburg.de}}
}
\vspace*{5mm}
\begin{center}
{\em Institut f\"ur Theoretische Physik und Astrophysik}\\
{\em Universit\"at W\"urzburg, D 97074 W\"urzburg, Germany} 
\end{center}

\vspace*{20mm}

\centerline{\bf Abstract}
\vspace*{2mm}
We consider discretized gravity in 4+2 dimensions compactified on a
disk of constant negative curvature. The curvature of the disk avoids the
presence of dangerous ultra-light scalar modes but comes also
along with a high multiplicity of states potentially
jeopardizing a good strong-coupling behavior of the discretized
theory. We demonstrate that for Standard Model matter propagating on
the five-dimensional boundary submanifold of the disk, the strong coupling scale, as seen by an observer,
can be parametrically larger than the local Planck scale. As a consequence,
we obtain a description of weakly coupled discretized gravity on the
boundary that can be compared with the continuum theory all the way up to the effective five-dimensional Planck scale.

\renewcommand{\thefootnote}{\arabic{footnote}}
\setcounter{footnote}{0}

\newpage

\section{Introduction}\label{sec:Introduction}
About a decade ago, the study of curved background geometries in more
than four dimensions has led to the development of important ideas
such as the Randall-Sundrum (RS) model \cite{Randall:1999ee,Randall:1999vf} and gauge gravity duality \cite{Maldacena:1997re}. Solving Einstein's equations for general curved backgrounds, however, is very difficult
in practice. Discretized or lattice gravity
\cite{Arkani-Hamed:2002sp}, on the other hand, may be a possibility
for describing gravity in extra-dimensional geometries at the level of
an effective field theory (EFT) \cite{Arkani-Hamed:2003vb,Schwartz:2003vj} (for
related work see, {\it e.g.}, \cite{Boulanger:2000rq}) without the
need to solve Einstein's equations explicitly
\cite{Randall:2005me}. In fact, it has been shown that
five-dimensional (5D) flat space lattice gravity gives a valid EFT of
massive gravity up to energies above the compactification scale
\cite{Arkani-Hamed:2003vb}, while discretized gravity in 5D warped space
works in the manifestly holographic regime \cite{Randall:2005me,Gallicchio:2005mh}.

Minimal models of 5D discretized gravity face, however, a strong coupling
phenomenon which forbids to take the continuum limit within a consistent
EFT. This difficulty arises from the presence of an ultra-light mode
with dangerous long-range interactions that prevent us from taking
the large volume limit \cite{Arkani-Hamed:2003vb}. In contrast to
this, the large volume limit can be taken in discretized 5D warped
space-time \cite{Randall:2005me,Gallicchio:2005mh}, but the EFT still
gets, due to the light mode, early strongly coupled below the local
Planck scale. A possible way out of this problem may be offered in
discrete hyperbolic space. If we view the 5th dimension as the
boundary of a six-dimensional (6D) hyperbolic disk, the light mode can
become very heavy due to the hyperbolic curvature of the disk
\cite{Bauer:2006pf}. This can leave us on the 5D boundary with a
theory of discretized gravity that is more weakly coupled than in
minimal 5D models. In fact, phenomenological features of hyperbolic extra
dimensions have been considered
previously in \cite{Kaloper:2000jb} and collider
implications were analyzed in
\cite{Bauer:2006pf,Melbeus:2008hk}. However, even on the hyperbolic
disk, the strong coupling scale as seen by a brane-localized observer
can become as large as that of the continuum theory only in the limit
where all massive gravitons decouple above the local Planck scale
\cite{Bauer:2006pf}.

In this paper, we consider a model for discretized gravity
on a 6D hyperbolic disk with an EFT for massive gravitons on the 5D
boundary that is meaningful at energies up to the effective 5D Planck scale. This is achieved by
letting  Standard Model (SM) matter propagate on the 5D boundary of the disk, just
as in universal extra dimensions \cite{Appelquist:2000nn}. Compared
with the previous case of an observer localized on a single point
\cite{Bauer:2006pf}, the propagation of SM matter on the boundary
implies KK number conservation, which shuts off the contribution of a high multiplicity of Kaluza-Klein (KK) levels to the tree-level
scattering amplitudes of SM matter. This gives a
theory of weakly coupled discretized gravity on the 5D boundary with a
strong coupling scale that is parametrically larger than the local
Planck scale. It thus becomes possible to compare the discretized
theory on the boundary with the continuum theory all the way up to the effective 5D Planck scale.

The paper is organized as follows. In Sec.~\ref{sec:geometry}, we
review the warped hyperbolic disk geometry. A rough discretization of
the disk for the flat and the strongly warped case is next carried out in
Sec.~\ref{sec:discretization}. Then, in Sec.~\ref{sec:strongcoupling},
we give for these cases the strong coupling scale in pure gravity. In
Sec.~\ref{sec:observer}, we consider the EFT of massive gravitons as
seen by an observer on the boundary of the disk. Finally, we
present in Sec.~\ref{sec:summary} our summary and conclusions.

\section{Disk geometry}\label{sec:geometry}
We start by briefly reviewing 6D general relativity compactified to four dimensions
on an orbifold $K_2/Z_2$, where $K_2$ is a two-dimensional disk of
constant negative curvature. We follow here closely the more detailed
description in \cite{Bauer:2006pf}. In what follows, the 6D coordinates $x^M$ are labeled by capital
Roman indices $M=0,1,2,3,5,6$ whereas Greek indices $\mu=0,1,2,3$ symbolize
the usual four-dimensional (4D) coordinates $x^{\mu}$ and the Minkowski metric is $\eta_{MN}=\textrm{diag}(1,-1,\dots,-1)$. On the disk $K_2$, a
point is given by polar coordinates $(r,\varphi)$, where
$r=x^{5}$ and $\varphi=x^{6}$ with $r\in[0,L]$ and
$\varphi\in[0,2\pi)$. In our notation, $r$ is
the geodesic distance of the point $(r,\varphi)$ from the
center and $L$ is the proper radius of the disk. We arrive at the orbifold
$K_2/Z_2$ by assuming the orbifold projection
$\varphi\rightarrow -\varphi$. This restricts the physical space on
the disk to $(r,\varphi)\in [0,L]\times [0,\pi]$.

The 6D metric $\tilde{g}_{MN}$ for the curved disk is defined by the line element
\begin{equation}
ds^{2}=e^{2\sigma(r)}g_{\mu\nu}(x,r,\varphi)dx^{\mu}dx^{\nu}-dr^{2}-\frac{1}{v^2}\,
\textrm{sinh}^2(v\cdot r)d\varphi^{2},\label{eq:6D-metric}\end{equation}
where $1/v$ is the curvature radius of the disk with $v>0$,
$g_{\mu\nu}(x,r,\varphi)$ is the 4D metric with $x\equiv(x^\mu)$, and
$\sigma(r)=-w\cdot r$, where $w$ is the curvature scale for the
warping. A comparison of our space with the Poincar\'e hyperbolic disk is given in
 \cite{Bauer:2006pf}. Like in RS I \cite{Randall:1999ee}, we have
 assumed in (\ref{eq:6D-metric}) orbifold boundary conditions with respect to $r$
at the center $r=0$ and at the boundary of the disk at $r=L$. While we identify
the UV brane with the center at $r=0$, we have IR branes residing at the orbifold fixed points on the boundary at
$r=L$.

We assume the Einstein-Hilbert action
\begin{equation}
\mathcal{S}=M_{6}^{4}\int d^{6}x\,\sqrt{|\tilde{g}|}(\tilde{R}-2\Lambda)\end{equation}
for gravity, where $\tilde{g}\equiv\det\tilde{g}_{MN}$, $M_{6}$ is the 6D Planck
scale, $\Lambda$ the bulk cosmological constant,
$\tilde{R}=\tilde{g}^{MN}\tilde{R}_{MN}$ the 6D curvature scalar, and
$\tilde{R}_{MN}$ the 6D Ricci tensor. Expanding $g_{\mu\nu}$ in terms of small fluctuations around 4D
Minkowski space as $g_{\mu\nu}=\eta_{\mu\nu}+h_{\mu\nu}$, we obtain,
to quadratic order, the linearized kinetic part of the graviton
Lagrangian density
\begin{equation}
\mathcal{S}_\text{lin}=\mathcal{S}_\text{kin}+\mathcal{S}_\text{FP}
\end{equation}
with the kinetic term
\begin{equation}\label{eq:kin}
\mathcal{S}_\textrm{kin}= M_6^4\int d^6x\,v^{-1}
\textrm{sinh}(vr)
\frac{1}{4}e^{2\sigma(r)}
(\partial^{\mu}h^{\nu\kappa}\partial_{\mu}h_{\nu\kappa}-\partial^{\mu}h\partial_{\mu}h-2h^{\mu}h_{\mu}+2h^{\mu}\partial_{\mu}h),
\end{equation}
where $h={h^{\mu}}_{\mu}, h_{\nu}=\partial^{\mu}h_{\mu\nu}$, and $\mathcal{S}_\text{FP}$ is of Fierz-Pauli form \cite{'tHooft:1974bx,Fierz:1939ix}\footnote{For an alternative to Fierz-Pauli mass terms see \cite{Dvali:2008em}.}
\begin{eqnarray}\label{eq:FP}
\mathcal{S}_\textrm{FP}&=& M_6^4\int d^6x\,v^{-1}
\textrm{sinh}(vr)\Big[
\frac{1}{4}e^{4\sigma(r)}
(\partial_{r}h_{\mu\nu})(\eta^{\mu\nu}\eta^{\alpha\beta}-\eta^{\mu\alpha}\eta^{\nu\beta})(\partial_rh_{\alpha\beta})\nonumber\\
&&\qquad\,\,+\frac{1}{4}e^{4\sigma(r)}v^2\textrm{sinh}^{-2}(vr)
(\partial_\varphi
h_{\mu\nu})(\eta^{\mu\nu}\eta^{\alpha\beta}-\eta^{\mu\alpha}\eta^{\nu\beta})(\partial_\varphi
h_{\alpha\beta})\Big].
\end{eqnarray}
As in \cite{Arkani-Hamed:2003vb}, we have neglected the 4D vector,
scalar, and radion degrees of freedom, and have assumed a gauge with $h_{5M}=h_{6M}=0$, for $M=0,1,\dots,6$.

\section{Warped space discretization}\label{sec:discretization}
Let us now consider the rough latticization of the disk $K_2$ as
described in \cite{Bauer:2006pf}. Different from earlier, however, we
now include a non-trivial warp factor that can be large. The latticization of the disk is
defined by a number sites and links, where $N$ sites, labeled as
$i=1,2,\dots,N$, are placed on the boundary and one site, carrying the
label $i=0$, is placed at the center ($r=0$) of the disk. The sites on the boundary of the disk are evenly spaced
on a concentric circle with proper radius $L$ around the central site
$i=0$, {\it i.e.} the $i$th site on the
boundary has polar coordinates $(r,\varphi)=(L,i\cdot\Delta\varphi)$, where
$\Delta\varphi=2\pi/N$. Two sites $i$ and $i+1$ on the boundary
are connected by a link $(i,i+1)_\textrm{link}$, while the site $i=0$ in the
center is connected to all $N$ sites on the boundary by links
$(0,i)_\textrm{link}$. To implement a lattice gravity theory in terms of this triangulation,
we will interpret the sites and links according to \cite{Arkani-Hamed:2002sp,Arkani-Hamed:2003vb,Schwartz:2003vj}.
Here, each site $i$ is equipped with its own metric
$g^{i}_{\mu\nu}$, which can be expanded around flat space as
$g^i_{\mu\nu}=\eta_{\mu\nu}+h^i_{\mu\nu}$, where $\eta_{\mu\nu}$ is
the usual 4D Minkowski metric.
In a naive latticization of the linearized action
$\mathcal{S}_\textrm{lin}$ in (\ref{eq:FP}), we then replace the derivatives on the sites as
\begin{equation}\label{eq:derivatives}
\partial_\varphi h_{\mu\nu}\rightarrow\frac{e^{-2wL}}{\Delta\varphi}
(h_{\mu\nu}^{i+1}-h_{\mu\nu}^i),\quad
\partial_r h_{\mu\nu}\rightarrow\frac{e^{-2wL}}{L}(h^i_{\mu\nu}-h_{\mu\nu}^0),\quad
M_6^4\int dx^5dx^6 \rightarrow M_4^2 \sum_{i=0,1}^N,
\end{equation}
where it is always understood that the summation starts
at $i=0$ for $\mathcal{S}_\text{kin}$ and at $i=1$ for
$\mathcal{S}_\text{FP}$. In the case of zero warping, $w=0$, the local
4D effective Planck scale $M_4$ on each of the $N+1$ sites is related to $M_6$ and
the usual 4D Planck scale $M_{\rm Pl}\simeq 10^{18}\:{\rm GeV}$ by
$M_4^2=M_6^{4}A/N$ and $M_{\rm Pl}^2=M_6^4A=M_4^2N$, where $A=4\pi v^{-2}\sinh^2{(vL/2)}$ is the proper area of the disk. For strong warping,
$w\gg v$ and $w\gg L^{-1}$, the graviton zero mode gets located at the center of
the disk and we have, instead, $M_4\approx M_\text{Pl}$. The model possesses the graviton mass spectrum
\begin{equation}\label{eq:eigenvalues}
M_{0}^{2}=0,\quad
M_{n}^{2}=m_{*}^{2}\epsilon^2+4m^{2}\epsilon^2{\textrm{sin}}^{2}\frac{\pi n}{N},\quad M_{N}^{2}=(N\epsilon^2+1)m_{\ast}^{2}\epsilon^2,
\end{equation}
where $m_{*}=1/L$ and $ m=Nv/2\pi\,\textrm{sinh}(vL)$ are the  proper inverse lattice spacings in radial ($m_{*}$) and
angular ($m$) direction, $n$ runs over $n=1,2,\dots,N-1$, and
$\epsilon\equiv e^{-wL}$ is the warp factor. In the basis
$(h_{\mu\nu}^{0},h_{\mu\nu}^{1},\dots,h_{\mu\nu}^{N})$, the
corresponding canonically normalized graviton mass eigenstates read
\begin{eqnarray}\label{eq:eigenstates}
M_0^2&:&\frac{1}{\sqrt{1+N\epsilon^2}}(1,\epsilon,\epsilon,\dots
,\epsilon)\hat{H}^0_{\mu\nu}(x),\nonumber\\
M_n^2&:&\frac{1}{\sqrt{N}}(0,1,e^{{\textrm{i}}\frac{2n\pi}{N}},e^{{\textrm{i}}\frac{4n\pi}{N}},\dots,e^{{\textrm{i}}\frac{2(N-1)n\pi}{N}})
\hat{H}^n_{\mu\nu}(x),\\
M_N^2&:&\frac{1}{\sqrt{N(1+N\epsilon^2)}}(N\epsilon,-1,-1,\dots, -1)\hat{H}^N_{\mu\nu}(x),
\nonumber
\end{eqnarray}
where we have, for simplicity, taken $N$ to be even. Note that
irrespective of the warping, the $N-1$ eigenstates with mass-squares
$M_n^2$ are all exactly located on the boundary of the disk. The
profiles of the zero mode and of the $N$th massive mode, however,
strongly depend on the choice of the warp factor. Consider first zero warping,
$\epsilon=1$. In this case, the zero mode has a flat wave function and
is mostly living on the boundary, whereas the heavy mode is peaked at
the center of the disk. In contrast to this, for strong warping,
$N\epsilon^2\ll1$, the mode with mass $M_N$ gets pushed away from the
center of the disk more towards the boundary. At the same time, we see
that the zero mode becomes, instead, strongly localized at the center of
the disk. We also observe that for $N\epsilon^2\ll 1$ a moderate
increase of $N$ has barely any effect on the gravitational
strength experienced at a single point on the boundary. This
reproduces the feature of RS that the low-energy Planck scale depends
only very weakly on the size of the warped extra dimension.

\section{Strong coupling scale in pure gravity}\label{sec:strongcoupling}
In this section, we discuss the strong coupling behavior of the
discretized model in the gravitational sector using the EFT for massive
gravitons in \cite{Arkani-Hamed:2002sp,Arkani-Hamed:2003vb,Schwartz:2003vj}.

To implement the EFT on the discretized disk $K_2$, we
replace in (\ref{eq:derivatives}) the differences
$h^i_{\mu\nu}-h^j_{\mu\nu}$ between the graviton fields on two sites $i$
and $j$ that are connected by a link $(i,j)_\textrm{link}$ according to
\begin{equation}\label{eq:restoringGCs}
h_{\mu\nu}^i-h_{\mu\nu}^j\,\,\rightarrow\,\,
g_{\mu\nu}^i-\partial_{\mu}Y_{i,j}^\alpha\partial_{\nu}Y_{i,j}^\beta
g^{j}_{\alpha\beta},\quad
Y_{i,j}^{\mu}(x_{\mu})=x^{\mu}+A_{i,j}^{\mu}(x_{\mu})+\partial^{\mu}\phi_{i,j}(x_{\mu}),
\end{equation}
in which $Y_{i,j}^{\mu}$ denotes a link field and $A_{i,j}^{\mu}$ and
$\phi_{i,j}$ are the vector and scalar components of the Goldstone bosons that
restore general coordinate invariances in the EFT (see \cite{Arkani-Hamed:2002sp,Arkani-Hamed:2003vb,Schwartz:2003vj}).

The kinetic Lagrangian of the gravitons is found by taking in (\ref{eq:restoringGCs}) the links
$(i,j)_\textrm{link}$ to be $(i,j)_\textrm{link}=(i,i-1)_\textrm{link}$, where
$i=2,\ldots ,N,N+1$, and $(i,j)_\text{link}=(i,0)_\text{link}$, where
$i=1,2,\dots,N,$ and $N+1\equiv 1$. As a consequence, we obtain from
$\mathcal{S}_\text{FP}$ in (\ref{eq:FP}), after partial integration, the total action of the disk in position space
\begin{eqnarray}\label{eq:Sdisk}
\mathcal{S}_{\rm disk}&=&\mathcal{S}_{\rm FP}+
M_6^4\int {\rm d}^6x\,
e^{-2wr}\big{(}h_{\mu\nu}\Box h_{\mu\nu}-m_*h_{\mu\nu}\Box\partial_r\phi-h_{\mu\nu}\Box\partial_\varphi\phi\nonumber\\
&&\qquad\qquad\qquad+e^{-2wr}m_*^2(\Box\phi)^3+e^{-2wr}m^2(\Box\phi)^3\big{)},
\end{eqnarray}
where we have used the short-hand notation
\begin{equation}
e^{-4wr}m_*^2(\Box\phi)^3\rightarrow e^{-4wr}m_*^2(\Box\phi_{0,i})^3,\quad
e^{-4wr}m^2(\Box\phi)^3\rightarrow e^{-4wr}m^2(\Box\phi_{i,i-1})^3.
\end{equation}
A linear Weyl-rescaling \cite{Arkani-Hamed:2003vb,Gallicchio:2005mh}
\begin{equation}\label{eq:Weylposition}
 h_{\mu\nu}\rightarrow h_{\mu\nu}'=h_{\mu\nu}+e^{-2wr}\eta_{\mu\nu}(m_*\partial_r+m^2\partial_\varphi)\phi
\end{equation}
removes the kinetic mixing between $h_{\mu\nu}$ and $\phi$ in
(\ref{eq:Sdisk}). In momentum space, for large $N$ and $m_\ast\gtrsim m$, the dominant
contribution to the massive graviton scattering amplitude is then given by the
tri-linear derivative coupling
\begin{equation}\label{eq:trilinear}
M_4^2\frac{m_*^2\epsilon^4}{\sqrt{N}}(\Box\Phi_n)(\Box\Phi_m)(\Box\Phi_{-n-m})
=\frac{1}{\sqrt{N}M_4m_*^4\epsilon^5}
(\Box\hat{\Phi}_n)(\Box\hat{\Phi}_m)(\Box\hat{\Phi}_{-n-m}),
\end{equation}
where we have expanded the scalar components of
the Goldstone fields belonging to the links $(i,0)_\textrm{link}$ as
$\phi_{i,0}=\frac{1}{\sqrt{N}}\sum_{n=1}^Ne^{{\rm i}2\pi i\cdot
n/N}\Phi_n$ with the definitions $\hat{\Phi}_{-n}=\hat{\Phi}_{N-n}$ and
$\hat{\Phi}_0=\hat{\Phi}_N$, introduced the canonically normalized fields as
$\hat{h}_{\mu\nu}=e^{-wr}h_{\mu\nu}$ and $\hat{\Phi}_n=\epsilon^3 M_4m_*^2\Phi_n$, and used $\epsilon=e^{-wL}$ for the
warp factor. From the amplitude for
$\hat{\Phi}_n-\hat{\Phi}_m$ scattering, we hence find for the discrete
disk geometry approximately the strong coupling scale
\begin{equation}\label{eq:strongcoupling}
\Lambda=(\sqrt{N}M_4m_*^4)^{1/5}\epsilon.
\end{equation}
For zero warping, the above equation becomes $\Lambda=(M_{\rm
  Pl}m_*^4)^{1/5}$, which is independent from the number of sites $N$ on the
boundary and identical with the strong coupling scale in the theory of
  a single massive graviton with mass $m_*$ \cite{Arkani-Hamed:2002sp}. Since all graviton modes (except for the zero mode)
  have received a mass of the order the inverse proper radius of the
  disk $m_*$, no dangerous light modes are present anymore and we take in
  (\ref{eq:strongcoupling}) the large $N$ limit. If, e.g., $N=100$ and
  $m_*\rightarrow M_4$, we find a strong coupling scale
  $\Lambda\approx 1.5\times\epsilon M_4$ that is somewhat
  larger than the (warped) local Planck scale on the boundary $\epsilon
  M_4$. If we wish to compare with the continuum theory, however, we
  cannot actually take the limit $m_*\rightarrow M_4$. The reason is
  that this would push all the graviton states above the local Planck
  scale $\epsilon M_4$ at which the continuum theory gets strongly coupled
  (cf.~(\ref{eq:eigenvalues})). When matching between the lattice and the
  continuum theory, we therefore always have to take $m_\ast<M_4$. The
  inverse lattice spacing in angular direction, $m$, on the other
  hand, can be made as large as $M_4$. The idea being that
  $m\rightarrow M_4$ practically represents a ``continuum limit'' of the boundary submanifold.

In estimating the strong coupling scales, we will always
assume that unwanted uneaten pseudo-Nambu-Goldstone bosons from the
mesh of links acquire large masses from invariant plaquette terms
that are added to the action \cite{Arkani-Hamed:2002sp} and lead to a
decoupling of these states. Moreover, the strong coupling scale in
(\ref{eq:strongcoupling}) has been estimated in pure gravity only.
In Sec.~\ref{sec:observer}, we will take the coupling to matter into
account and determine the strong coupling scale that is actually seen
by an observer made of SM matter propagating on the 5D boundary submanifold of the disk.

\section{Coupling to matter}\label{sec:observer}
Let us next determine the observed strong coupling scale for SM matter
propagating on the boundary of the disk. Previously, we have studied the
case where the SM is located on a single point on the boundary of the disk
\cite{Bauer:2006pf}. In this case, the local interaction ensures that
the KK scalar components of the Goldstones couple all with equal strength
to matter. Due to the high multiplicity of these states, this leads to
an observed strong coupling scale that can be at most as large as the
local Planck scale $\epsilon M_4$.

Different from this scenario, we will now assume that the SM matter
propagates as zero modes on the 5D boundary $(r,\varphi)=(L,\varphi)$ of
the disk as in the universal extra dimension scenario. From
(\ref{eq:eigenstates}), we find that SM matter will then interact on
the boundary of the disk with gravity as (see also \cite{Davoudiasl:1999jd})
\begin{equation}\label{eq:interaction}
\mathcal{S}_{\rm int} \approx \frac{1}{M_\text{Pl}}\int d^{4}x\,dy\,\frac{1}{R}
T^{\mu\nu}\Big(\hat{H}^{0}_{\mu\nu}
+
\frac{1}{\epsilon}
\sqrt{\frac{1+N\epsilon^2}{N}}
\sum_ne^{-{\rm i} n\cdot y/R}\hat{H}^{n}_{\mu\nu}
-\frac{1}{\epsilon\sqrt{N}}\hat{H}^{N}_{\mu\nu}\Big)+{\rm
h.c.},
\end{equation}
where $T^{\mu\nu}$ is the SM stress-energy tensor, $y\rightarrow
i\cdot a_\varphi$ ($i=1,2,\dots$) is the coordinate of the 5D
submanifold on the boundary, $dy\rightarrow a_\varphi$, and $R$ is the
proper length of the boundary. On the disk $K_2$, $n$ would run from
$1$ to $N-1$, but after orbifolding we have $n=1,\dots,\frac{N}{2}-1$
(for $N$ even). We will work here, for simplicity, with the states
found in the discretization of the disk $K_2$ in
Sec.~\ref{sec:discretization}, with the understanding that we actually
have to finally impose orbifold boundary conditions as explained in
Sec.~\ref{sec:geometry} for the continuum theory.\footnote{For a
  discussion of differences between lattice gravity on an interval and
  on a circle see \cite{Arkani-Hamed:2003vb}.}

We see from (\ref{eq:interaction}) that SM matter couples with gravitational strength
$M_\textrm{Pl}^{-1}$ to the graviton zero mode $\hat{H}^0_{\mu\nu}$ and to the heavy modes
$\hat{H}^n_{\mu\nu}$ roughly with strengths $\lesssim(\epsilon
M_\text{Pl})^{-1}$ (for small $N$). For $m_*^2\gtrsim m$, the Weyl rescaling in
(\ref{eq:Weylposition}) corresponds in momentum basis approximately to
the change of the canonically normalized fields
\begin{equation}\label{eq:Weylmomentum}
\hat{H}^n_{\mu\nu}\rightarrow\hat{H}^n_{\mu\nu}+\eta_{\mu\nu}\hat{\Phi}_{-n},
\end{equation}
where $n=1,\dots,N$, while $\hat{H}^0_{\mu\nu}$ has no kinetic mixing
with the Goldstones. The Weyl rescaling in (\ref{eq:Weylmomentum}) introduces in
$\mathcal{S}_\text{int}$ the matter-Goldstone interactions
\begin{equation}
\mathcal{S}^\Phi_\text{int}\approx \frac{1}{\epsilon M_\text{Pl}}\int
d^4x\,dy\frac{1}{R}T\Big(
\sqrt{\frac{1+N\epsilon^2}{N}}
\sum_n e^{-{\rm i}n\cdot y/R}\hat{\Phi}_{-n}
+\frac{1}{\sqrt{N}}\hat{\Phi}_0\Big)+{\rm h.c.},
\end{equation}
where $T=\text{Tr}(T_{\mu\nu})$ and $\hat{\Phi}_0=\hat{\Phi}_N$ (see Sec.~\ref{sec:strongcoupling}). Integrating now over $y$, we obtain in the
4D effective low-energy theory the interaction Lagrangian
\begin{equation}\label{eq:4Dinteractions}
\mathcal{S}^\Phi_\text{int}\approx \frac{1}{\epsilon M_\text{Pl}}\int
d^4x\, T\sum_n\mathcal{F}^{n}\hat{\Phi}_{-n}+ {\rm h.c.},
\end{equation}
where $\mathcal{F}^n$ is a form factor and $n$ starts now from zero. The important point is that
for SM matter propagating on the boundary, KK number conservation
implies $\mathcal{F}^0\sim 1/\sqrt{N}$ but $\mathcal{F}^n=0$ for $n\neq 0$. If
the SM fields were, instead, localized on a single point
on the boundary \cite{Bauer:2006pf}, we would have
$\mathcal{F}^n\sim 1$ for all $n$, i.e.~a universal coupling of SM matter to a
high multiplicity of Goldstones.

Let us next denote by $\psi^n$ the KK resonances of some SM field. The
interactions of $\psi_i$ and $\psi_j$ with the Goldstones
$\hat{\Phi}_{-n}$ are then described by (\ref{eq:4Dinteractions}),
with the replacements $T\rightarrow T^{ij}$ and
$\mathcal{F}^n\rightarrow \mathcal{F}^{ijn}$, where $T^{ij}$ is the
trace of the energy-stress tensor for the fields $\psi_i$ and $\psi_j$, while
$\mathcal{F}^{ijn}$ is the corresponding form factor describing the
coupling of $T^{ij}$ with $\hat{\Phi}_{-n}$. For the zero modes of the
matter fields $i=j=0$ we have $T^{00}=T$ and $\mathcal{F}^{00n}=\mathcal{F}^n$.
Feynman rules for this case with gravity and matter on the boundary
can be derived as in \cite{Macesanu:2003jx}
(cf.~\cite{Han:1998sg}). Again, KK number conservation will give
$\mathcal{F}^{ijn}\sim\sqrt{(1+N\epsilon^2)/N}$ only for $i+j=n$, whereas
$\mathcal{F}^{ijn}=0$ otherwise.

\begin{figure}
\begin{center}
\includegraphics*[bb = 70 610 539 730]{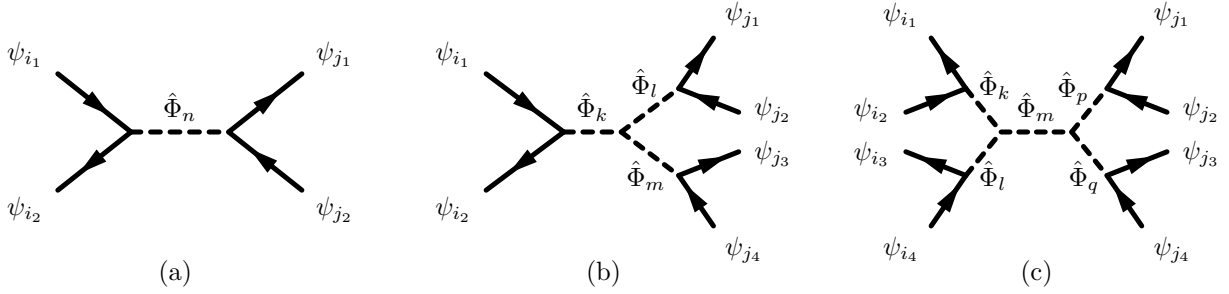}
\end{center}
\vspace*{-5mm}
\caption{\small Scattering of KK fermions via Goldstone modes at tree level. KK
  number conservation at the matter-Goldstone vertices switches off
  the high multiplicity of internal states.}\label{fig:matterscattering}
\end{figure}

We are now in a position to estimate the strong coupling scale
associated with processes involving SM KK resonances $\psi_i$ and
Goldstones $\hat{\Phi}_n$. At tree level, we will restrict to the three
example processes shown in  Fig.~\ref{fig:matterscattering}: (a)
$\psi_{i_1},\psi_{i_2}\rightarrow\psi_{j_1},\psi_{j_2}$, (b)
$\psi_{i_1},\psi_{i_2}\rightarrow \psi_{j_1},\dots,\psi_{j_4}$, and
(c) $\psi_{i_1},\dots,\psi_{i_4}\rightarrow\psi_{j_1},\dots,\psi_{j_4}$.
In what follows, we assume that the $\psi_i$ are fermions. Let $E$ be the typical energy of the external particles. The
largest matter-Goldstone vertex factors are those involving
$\hat{\Phi}_{n\neq 0}$ and small $N$, which are of the order
$E/\epsilon M_\text{Pl}$. Each three-Goldstone vertex, on the other
hand, contributes to a diagram a factor
$E^6/\sqrt{N}M_4m_*^4\epsilon^5$. We will perform a conservative
estimate and assume for the matter-Goldstone vertices this
limit. One should, however, keep in mind that the actual couplings of
the SM fermion states to $\hat{\Phi}_0$ are smaller and suppressed by a factor
$1/\sqrt{N}$ with respect to the couplings that we will use below.
We then obtain for the scattering amplitude $\mathcal{A}$ of diagram (a) roughly
\begin{equation}
\mathcal{A}\sim\left(\frac{E}{\epsilon M_\text{Pl}}\right)^2\frac{1}{E^2}=
\frac{1}{\epsilon^2M_\text{Pl}^2}.
\end{equation}
From (\ref{eq:trilinear}), we find that the amplitude for diagram (b) is given by
\begin{equation}
\mathcal{A}\sim\left(\frac{E}{\epsilon
  M_\text{Pl}}\right)^3\left(\frac{1}{E^2}\right)^3\frac{E^6}{\sqrt{N}M_4
  m_*^4\epsilon^8}=\frac{E^3}{\sqrt{N}M_\text{Pl}^3 M_4m_*^4\epsilon^8},
\end{equation}
where we have used momentum conservation. Similarly, we arrive for
diagram (c) at the amplitude
\begin{equation}
\mathcal{A}\sim
\left(\frac{E}{\epsilon M_\text{Pl}}\right)^4\frac{E^{12}}{N
  M_4^2m_*^8\epsilon^{10}}\left(\frac{1}{E^2}\right)^5=\frac{E^6}{NM_\text{Pl}^4M_4^2m_*^8\epsilon^{14}}.
\end{equation}
Note that KK number conservation ensures that the amplitudes are independent from $N$. This is completely different from
\cite{Bauer:2006pf}, where, for the SM located at a single point on
the boundary of the disk, the amplitudes corresponding to the diagrams
(a), (b), and (c), would scale as $N,N^2,$ and $N^3$, respectively. We thus estimate that the strong coupling scales associated with the diagrams in Fig.~\ref{fig:matterscattering} become approximately
\begin{equation}\label{eq:strongcouplings}
\text{(a)}\::\:\Lambda\rightarrow\epsilon M_\text{Pl},\qquad
\text{(b)}\::\:\Lambda\rightarrow(\sqrt{N}M_\text{Pl}^3 M_4m_\ast^4)^\frac{1}{8}\epsilon,\qquad
\text{(c)}\::\:\Lambda\rightarrow(\sqrt{N}M_\text{Pl}^2M_4m_*^4)^\frac{1}{7}\epsilon.
\end{equation}
In the limit $m_*\rightarrow M_4$, the strong coupling scale
$\Lambda$ becomes for both zero ($\epsilon=1$) and strong warping
($\epsilon\ll 1$) larger than the local Planck scale $M_4$. 

\begin{figure}
\begin{center}
\includegraphics*[bb = 130 610 530 710]{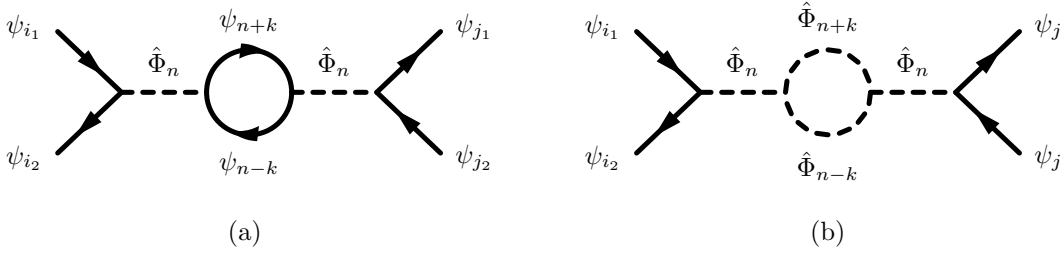}
\end{center}
\vspace*{-5mm}
\caption{\small One loop corrections to the amplitude in
  Fig.~\ref{fig:matterscattering} (a). KK number conservation requires
  $i_1+i_2=j_1+j_2=n$.}\label{fig:loops}
\end{figure}

Let us next look at quantum corrections to the scattering
amplitudes. Consider for this purpose the 1-loop diagrams (a) and (b) in Fig.~\ref{fig:loops}.
The diagrams give rise to the loop expansion parameters
\begin{equation}\label{eq:loopexpansions}
\text{(a)}\::\:\epsilon_1=\frac{1}{16\pi^2}\left(\frac{E}{M_\text{Pl}}\right)^2N_\text{KK}(E),\qquad\text{(b)}\::\:\epsilon_2=\frac{1}{16\pi^2}\left(\frac{E^5}{\sqrt{N}M_4m_*^4\epsilon^5}\right)^2N_\text{KK}(E),
\end{equation}
where $N_\text{KK}(E)$ is the number of KK states below $E$. The
boundary has a warped compactification scale $M_c=1/R=2\epsilon
m/N$. With $N_\text{KK}(E)=ER$, we find that the loop expansion parameters in
(\ref{eq:loopexpansions}) imply for zero warping ($\epsilon=1$) the strong coupling scales
\begin{subequations}\label{eq:loops}
\begin{equation}\label{eq:flatloop}
\text{(a)}\::\: \Lambda\rightarrow(16\pi^2)^\frac{1}{3}M_4,\qquad
\text{(b)}\::\:\Lambda\rightarrow(16\pi^2)^\frac{1}{11}(M_4^3m_*^8)^\frac{1}{11}\qquad(\epsilon=1)
\end{equation}
where we have used the 5D flat space relations
$M_4=(M_\text{Pl}^2/R)^\frac{1}{3}=M_\text{Pl}/\sqrt{N}$. Likewise,
in the strongly warped case ($N\epsilon^2\ll 1$), it follows
from (\ref{eq:loops}) that the theory would get strong coupled at
\begin{equation}\label{eq:warpedloop}
\text{(a)}\::\: \Lambda\rightarrow(16\pi^2)^\frac{1}{3}(M_\text{Pl}^2/R)^\frac{1}{3},\qquad
\text{(b)}\::\:\Lambda\rightarrow(16\pi^2)^\frac{1}{11}(M_\text{Pl}^2m_*^8m)^\frac{1}{11}\epsilon\qquad(\epsilon\ll1)
\end{equation}
\end{subequations}
where we have applied the warped space relation $M_4\approx M_\text{Pl}$. For $m_*,m\rightarrow M_4$, diagram (b) then leads
for a general warp factor to a strong coupling scale
\begin{equation}
 \Lambda\approx 1.5\times \epsilon M_4,
\end{equation}
which is larger than the local Planck scale $\epsilon M_4$. This is similar to the scale that follows from
Goldstone boson scattering in (\ref{eq:strongcoupling}) in the same
limit. Again, we cannot really take $m_*\rightarrow M_4$, since
this would lead to a decoupling of all massive graviton states above
the local Planck scale. But having $m_*$ somewhat smaller
than $M_4$, allows us to compare the lattice gravity model with the
continuum theory all the way up to the warped strong coupling scale $\epsilon
M_4$ of the effective 5D continuum theory on the boundary.

\begin{figure}
\begin{center}
\includegraphics*[height=5cm]{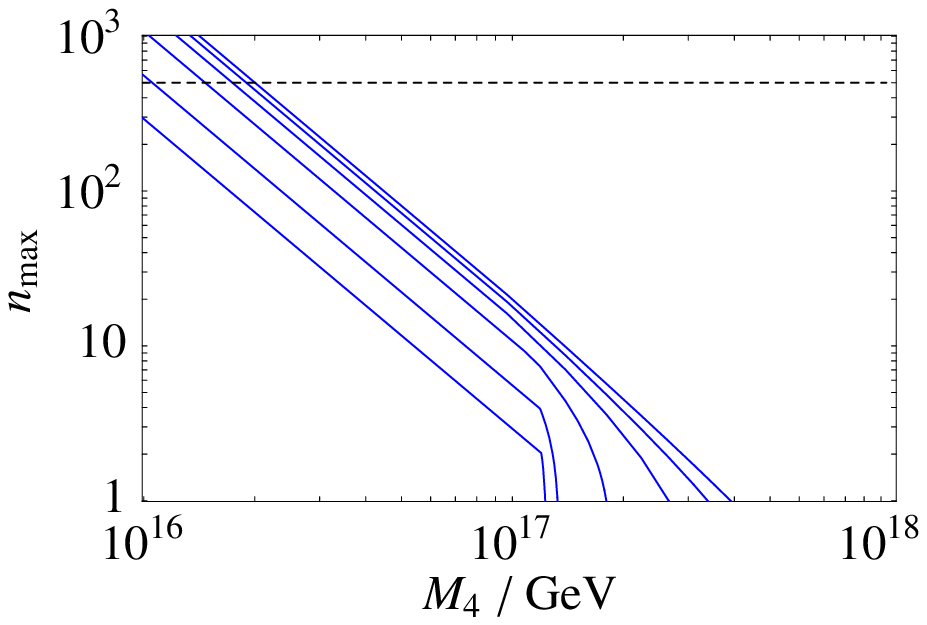}
\includegraphics*[height=5cm]{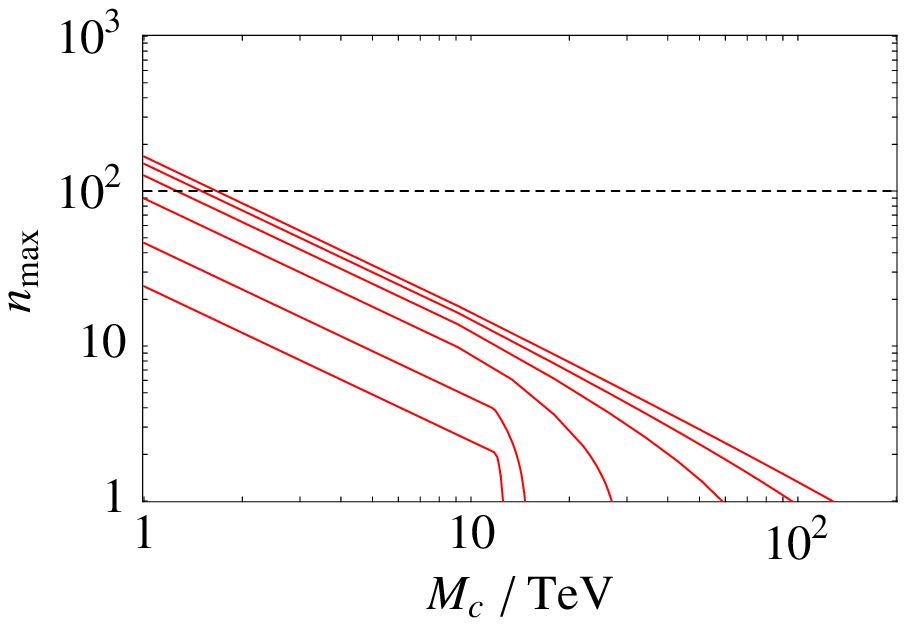}
\end{center}
\vspace*{-5mm}
\caption{\small Maximum number of KK gravitons below the fundamental
  scale $M_4$ (see text) for the flat (left panel) and warped case
  (right panel). In the warped case, we have assumed a warp factor $\epsilon=10^{-12}$.}\label{fig:nmax}
\end{figure}

Let us see how many KK gravitons we can actually fit in below $\epsilon
M_4$. In Fig.~\ref{fig:nmax}, we show the maximum number
$n_\text{max}$ of KK gravitons with masses below the 5D effective
Planck scale on the boundary, $\epsilon M_4$, for both the flat and
the strongly warped case. For nonzero warping, we have assumed a warp factor $\epsilon=10^{-12}$, leading to a local Planck scale on the
boundary of the order $10^3\,\text{TeV}$. In Fig.~\ref{fig:nmax}, the
left panel depicts the $n_\text{max}$ as a function of the local
Planck scale $M_4$ and the right panel as a function
of the warped compactification scale $M_c\simeq 1/R$ of the 5D
boundary. The dashed lines in the figures indicate that above
$10^2-10^3$ KK states the theory with the SM in universal extra
dimensions would become non-perturbative at the one-loop level
\cite{Appelquist:2000nn,Nomura:2001tn}, i.e.~we have always to restrict in
any case to $n_\text{max}\lesssim 10^2-10^3$ (around a TeV, $n_\text{max}$ it is
closer to $10^2$ because of the running of the QCD coupling constant). In both graphs, we have
always set $m=M_4$ to practically achieve the ``continuum
limit'' on the boundary in the sense of Sec.~\ref{sec:strongcoupling}. The curvature scale of the disk $m_*<M_4$,
has been chosen in such a way that the smaller of the
two strong coupling scales estimated in (\ref{eq:strongcoupling}) and
case (b) in (\ref{eq:loops}) becomes equal to some scale $a\cdot
M_4>M_4$, where $a$ is a positive number larger than one. In
Fig.~\ref{fig:nmax}, $a$ takes for both graphs from top to bottom in
this order the values $a=1.0,1.15,1.30,1.45,1.55,1.58$. One can see
that the larger $a$, the smaller $n_\text{max}$, corresponding to an
increasing number of KK states that decouple above $\epsilon M_4$, although this
difference becomes marginal for $n_\text{max}\gg 1$. Having a number
of about $n_\text{max}=\mathcal{O}(10^2)$ KK states below the strong
coupling scale requires therefore in the flat case the local Planck
scale to be in the range
\begin{equation}\label{eq:flatparameters}
10^{16}\,\text{GeV}\lesssim M_4\lesssim
10^{17}\,\text{GeV}\quad(\epsilon=1).
\end{equation}
For $n_\text{max}\gtrsim\mathcal{O}(10)$, the strong coupling scale
set by the loop in diagram (b) in Fig.~\ref{fig:loops} dominates by
dropping below that given in (\ref{eq:strongcoupling}) and scales like $n_\text{max}\sim\epsilon
M_4/M_c$. Therefore, going for the warped case in Fig.~\ref{fig:nmax}
to warp factors $\epsilon<10^{-13}$ will allow only a few states
$n_\text{max}\lesssim \mathcal{O}(10)$ below $\epsilon M_4$. A
perturbative model with about $\sim 10^2$ states below the local Planck
scale $\epsilon M_4$ therefore requires warp factors and warped boundary
compactification scales of the orders
\begin{equation}\label{eq:warpedparameters}
 \epsilon\gtrsim 10^{-12},\quad M_c\gtrsim 10\,\text{TeV}.
\end{equation}
To solve the gauge hierarchy problem, one may therefore have
to resort to supersymmetry. In total, we observe for the ranges of parameters in
(\ref{eq:flatparameters}) and (\ref{eq:warpedparameters}) that we can have
a number of about $\mathcal{O}(10^2)$ KK graviton states with a strong
coupling scale as large as $\epsilon M_4$. This
allows to compare the lattice theory with the effective 5D continuum
theory on the boundary up to the effective warped 5D Planck scale $\epsilon M_4$.

\section{Summary and conclusions}\label{sec:summary}
In this paper, we have studied discretized gravity in a 6D
geometry, where the two extra dimensions have been compactified on a hyperbolic
disk. We have studied for the warped case the extreme limit of only two
(or three) lattice sites in radial direction but with a smooth
discretization and many sites on the boundary.

Working on the hyperbolic disk has the advantage that light
states endangering a good strong coupling behavior of the discretized
theory can be made massive by switching on the hyperbolic
curvature. We have estimated the strong coupling scale of discretized as seen by
an observer made of SM matter propagating on the 5D boundary of the
disk. We found that the observed strong coupling scale can be parametrically larger than the Planck scale of the 5D
effective continuum theory. This is due to KK number conservation which
shuts off the high multiplicity of KK states that would otherwise
contribute to tree level scattering amplitudes for an observer
localized at a single point. We have estimated the range of parameters
necessary for a theory of about $\sim 10^2$ massive
gravitons in discretized gravity to be perturbative up to the
effective 5D Planck scale for general warping. For these parameters, the
discretized gravity theory on the boundary admits a comparison with the
continuum theory all the way up to the strong coupling scale of the
effective 5D continuum theory.

It would be interesting, e.g., to see how our results could be used for the
formulation of a weakly coupled discrete version of the usual RS model
as an effective theory, to compare them to theories dual to
large-$N$ quantum field theories \cite{Kiritsis:2008at}, and to apply
them to modifications of gravity with relation to cosmology \cite{Dvali:2000rv,Ohl:2008tw}.

\section*{Acknowledgements}
I would like to thank T.~Ohl for valuable discussions. This work was supported by the Federal Ministry of Education and Research (BMBF) under contract number 05HT6WWA.

\end{document}